\documentclass[prd,superscriptaddress,showpacs,preprintnumbers,amsmath,amssymb]{revtex4}

\usepackage[latin1]{inputenc}
\usepackage{graphicx}
\usepackage{amsmath}
\usepackage{float}
\usepackage{color}
\usepackage{bm}
\usepackage{amssymb}

\def\be{\begin{equation}}
\def\ee{\end{equation}}
\def\bea{\begin{eqnarray}}

\def\eea{\end{eqnarray}}
\def\beq{\begin{eqnarray}}
\def\eeq{\end{eqnarray}}

\begin{document}
%%%%%%%%%%%%%%%%%%%%

\title{Inflation with improved D3-brane potential and the fine tunings \\associated with the model }
%______________________________________
\author{Amna Ali}
\affiliation{Centre of Theoretical Physics, Jamia Millia Islamia,
New Delhi-110025, India}

\author{Atri Deshamukhya}
\affiliation{Department of Physics, Assam University, Silchar,
Assam-788011, India}
\author{Sudhakar Panda}
\affiliation{Harish-Chandra Research Institute, Chhatnag Road,
Jhusi, Allahabad-211019, India}

\author{M.~Sami\footnote{Senior Associate, Abdus Salam International Centre for Theoretical Physics, Trieste, Italy.}}
\affiliation{Centre of Theoretical Physics, Jamia Millia Islamia,
New Delhi-110025, India}

\begin{abstract}
We revisit  our earlier investigations of the brane-antibrane
inflation in a warped deformed conifold background, reported in
\cite{ACPS}, where now we
 include the contributions to the inflation potential arising from imaginary anti-self-dual (IASD) fluxes including
 the term with irrational scaling dimension discovered recently in\cite{BDKKM12}.
 We observe that these corrections to the effective
potential help in relaxing the severe fine tunings associated with
the earlier analysis. Required number of e-folds, observational
constraint on COBE normalization  and low value of the tensor to
scalar ratio  are achieved which is consistent with WMAP seven years
data.

\end{abstract}

\maketitle

\section{Introduction}
%%%%%%%%%%%%%%%%%
 Cosmological inflation \cite{inflation} is a mechanism, for the universe to undergo a brief period of accelerated expansion. This is postulated to
  cure some of the intrinsic problems of the hot big-bang model. The
inflationary scenario not only explains the large-scale homogeneity
of our universe but also provides a way, via quantum fluctuation, to
generate the primordial inhomogeneities which is the seed for
understanding the structure formation in the universe. Such
inhomogeities have been observed as anisotropies in the temperature
of the cosmic microwave background.The inflationary paradigm has
stood the test of observational and theoretical challenged in past
two decades\cite{Spergel1,Spergel2}.

Inflation can be implemented using a single scalar field slowly
rolling over the slope of its potential. However, the initial
conditions for inflation and the form of the potential function are
expected to come from a fundamental  theory of gravity and not to be
chosen arbitrarily. In this context, enormous amount of efforts are
underway to derive inflationary models from string theory, a
consistent quantum field theory  around the Planck's scale and
considered to be ultraviolet-complete theory of gravity . Discovery
of D-branes, gauge/gravity duality and various nonperturbative
aspects in string theory have also played crucial role in  building
and testing inflationary models of cosmology.

In past few years, many inflationary models have been constructed from string theory compactified to four dimensions where D-branes have played the most important role.  The examples are inflation due to
tachyon condensation on a non-BPS brane, inflation due to the motion
of a D3-brane towards an anti-D3-brane
\cite{sen,linde,kallosh,lindeD}, inflation due to geometric tachyon
arising from the motion of a probe brane in the background of a
stack of either NS5-branes or the dual D5-branes \cite{GTach} . Another attempt has been made in \cite{eliz} where-higher curvature corrections and a dynamical dilaton has been included in the effective action for generating dark energy.
However,  these models do not take into account the details of
compactification and the effects of moduli stabilization. Such issues could be addressed only when it was
learnt \cite{GKP} that the background fluxes sourced by D3 and D5-branes
can stabilize the axio-dilaton and the complex structure moduli of
type IIB string theory compactified on an orientifold of a
Calabi-Yau threefold. Moreover, the back reaction of these D-branes yields the geometry of a throat \cite{KS} which could be glued smoothly to the compact Calabi -Yau manidold.Further important progress was achieved when
it was shown in Ref.~\cite{KKLT} that the K\"ahler moduli fields
also can be stabilized by a combination of fluxes and
nonperturbative effects via gauge dynamics of either an Euclidean D3-brane
or from a stack of   D7-branes, wrapping super-symmetrically a
four cycle of the compact manifold, placed around the base of the throat.

The above results enabled to construct an inflationary model \cite{KKLMMT} which took  account of the compactification data (see
also Refs.~\cite{dbpapers}) since the inflaton potential is obtained by
performing string theoretic computations involving the details of
fluxes and warping. In this scenario inflation is realized by
the motion of a D3-brane, placed in the compact manifold, towards a distant static anti-D3-brane
sitting at the tip of the throat. The radial separation between
the two is considered to be the inflaton field. The effect of the
moduli stabilization resulted in a large mass to the inflaton field
which turned out to be of the order of Hubble parameter and
hence spoils the inflation.  As a possibility for circumventing this problem, it was proposed
\cite{Bau2,Bau3,KP} to embed  the D7-branes such that at least one of the
four-cycles carrying the nonperturbative effects descend down a
finite distance into the warped throat which implied that the probe D3-brane is
constrained to move only inside the throat. This consideration lead to the inflaton potential having an inflection point. The inflation dynamics however, within a single throat model \cite{PSS},
 revealed that when the spectral index of scalar perturbation reaches the scale
invariant value, the amplitude tends to be larger than the COBE
normalized value by about three order of magnitude, making the model
unrealistic. Such  problems could  be solved if another non-inflating throat is added to the compactification procedure \cite{HC}.

\section{Potential from compactification effects and its implication for Inflation}

The possibility of a realistic model for
brane inflation, within the large volume compactification
and a single throat scenario, received further attention when the authors of Ref.~\cite{BDKKM}
observed that there are corrections to the inflaton potential
which arise from the compactification effects in the ultra violet (UV). Thus the assumption of the D7-brane
 descending into the throat was relaxed. Instead, using AdS/CFT correspondence, the throat geometry  (conifold ) has
  been treated as an approximate conformal field theory with a high cut off scale $M_{UV}$.  In this context, the position
  of the probe D3-brane is identified with the Coulomb branch vev of a field in the gauge theory which couples to  bulk moduli fields.
  This coupling changes the K\"ahler potential and hence the inflaton potential. It was noted that the leading contribution
  to the inflaton potential comes from the coupling of a chiral operator with dimension $\Delta = 3/2$ in the dual gauge
   theory to the bulk field. Taking this contribution in to account, the full inflaton potential takes the form:
\be V(\phi)~=~D\left[ 1 + \frac{1}{3} \left(
\frac{\phi}{M_{pl}} \right)^2 - C_{3/2}
\left(\frac{\phi}{M_{UV}}\right)^{3/2}  - \frac{3D}{16\pi^2 \phi^4} \right] \label{Spot} \ee

where $\phi$ is the canonically normalized inflaton field related to the position of the D3-brane; $C_{3/2}$ is a positive
constant and $D \sim 2 a_0^4 T_3$. Here $a_0$ is the minimal warp factor at the tip of the throat and $T_3$ is the tension of
D3-brane. Note that the third term in the above equation is the contribution coming from the compactification effect in the
 UV and the rest of the potential is the same as in \cite{KKLMMT}.

The inflationary dynamics, using the above potential, was investigated in \cite{ACPS} and the reheating issue was discussed in \cite{PST}.
It was observed that the parameter $C_{3/2}$ in the above potential has to be severely fine tuned for the inflation model to be
consistent with the WMAP five years data \cite{KAM1}. This fine tuning becomes worse for the consistency with recent WMAP seven years data\cite{KAM2}.
Moreover, the initial condition for the inflaton field value has also to be chosen, with acute fine tuning, to yield a possible maximum fiftyeight  e-foldings.

On the other hand, recently the authors of Refs.~\cite{BDKKM12} have
performed a detailed analysis of the potential on the Coulomb branch
of the conifold gauge theory taking into account of the UV effect on
the probe D3 brane and have found many more corrections to the
inflaton potential. The general structure of the corrections arising
from UV deformations of the background take the form (See
\cite{BDKKM12} for details):

\be
V_c (\phi) = \sum_i  C_i \frac{\phi^{\Delta_i}}{M_{UV}^{\Delta_i - 4}}
\ee
where $M_{UV}$ is a UV mass scale related to the ultraviolet location at which
the throat is glued into the compact bulk. While, the constant coefficients $C_i$
are left undetermined, the scaling dimensions $\Delta_i$ are found to be $1, 3/2, 2, 5/2, L=2\sqrt{7}-5/2, \cdots$.

It should be noted that the parameters $C_i$ are not directly
restricted by flux compactification and Gauge/gravity
correspondence. These coefficients contain information from two
sources:(1) product of trigonometric function involving the five
angles of $T^{(1,1)}$ space (the radial component being the
inflaton, $\phi$ used in the paper). So these are objects whose
numerical values are less than one, (2) they involve small
perturbations of fields in the gauge theory side, which like in any
field theory are much less than one and are not apriori fixed but
can be fixed only with some dynamical process where we have some
information from observation. In fact the parameters $C_i$ are
product of these two small values. Thus the only restriction we have
is that they are much much less than one.
%%%

These contributions arise from various sources. For example, terms with scaling
dimensions $\Delta_i$ with $i = 3/2, 2, \cdots$ come from homogeneous solution
  (an arbitrary harmonic function on the conifold) and with $i = 1, 2, 5/2, L=2\sqrt{7}-5/2, \cdots$
  come from inhomogeneous solutions sourced by fluxes.   The term with scaling dimension $2\sqrt{7}-5/2$
  corresponds to a flux perturbation dual to a non-chiral operator which is generically present
   but is not captured via perturbations of the superpotential.
%%%

In what follows, we  reinvestigate the inflation dynamics by
including these new corrections with the hope that the problems as
mentioned earlier can be cured namely to examine the flexibility in
the space of parameters to obtain the required number
 of e-foldings and other consistency of the inflation dynamics. In fact  we find that these corrections not only help in constructing a
  viable inflation model consistent with WMAP seven years data but also the fine tuning problem as mentioned above is considerably relaxed.
  We can easily obtain sixty e-foldings and even more without being too tight in the choice of initial conditions and numerical values of
  the parameters. For this purpose, we write the full inflaton potential, including $V_c$ as given above, as follows:

\begin{equation}
{\cal
V} = {\cal D}\left[1 - C_1 x - C_{3/2}x^{3/2} + (\frac{\alpha^2}{3} - C_2) x^2 - C_{5/2} x^{5/2} - C_L x^L - \frac{3{\cal D}}{16 \pi^2\alpha^4 x^4}\right]
\label{Spot2}
\end{equation}
where $x=\phi/M_{UV}$, ${\cal V}=V/M_{UV}^4$,  ${\cal D}=D/M_{UV}^4$ and  $\alpha=M_{UV}/M_{pl}$ and  $L=2\sqrt{7}-5/2$  .
%%%%%%%%%%
Let us note that as we are dealing with a small field inflation
model, we only retain terms up to ${\Delta = 2\sqrt{7}-5/2}$ as other
contributions are insignificant. In what follows, we shall consider
the inflationary dynamics of the field with the modified potential
given by (\ref{Spot2}).

\section{Slow roll inflation}
In this section, we shall study the dynamics of inflation based upon
the improved D-brane potential and  demonstrate that the model under
consideration based upon (\ref{Spot2}) performs much better than the
earlier model given by (1). The modified potential allows to easily
generate enough inflation without involving the fine tuning of
initial conditions for slowly rolling inflaton.  The corrected
potential also allows to decrease the fine tuning of model
parameters present in   (\ref{Spot}) in a significant fashion, as we
show below.

For the sake of convenience, let us
 cast the evolution equation for the field ,
$\ddot{\phi}+3 H\dot{\phi}+V_{,\phi}=0  $ and the Friedmann equation
$H^2=\left(\frac{\dot{\phi}^2}{2}+V(\phi)\right)/3 M_{pl}^2
$
in the  autonomous form
\begin{eqnarray}
&&\frac{dx}{dN}=\frac{y}{{\cal H}} \\
&&\frac{dy}{dN}=-3y-\frac{1}{{\cal H}}\frac{ d{\cal V}}{dx}\\
&&{\cal H}^2=\frac{\alpha^2}{3}\left(\frac{1}{2}{y}^2+{\cal
V}(x)\right)
\end{eqnarray}
where $y=\dot{x}/M_{UV}^{2}$, ${\cal H}=H/M_{UV}$,   and $N$ designates the number of e-foldings.
  In the scenario under consideration, the mobile
D3-brane  moves towards the anti-D3-brane located at the tip of the
throat corresponding to $x=0$ and thus we have  $0< x < 1$ since
$\phi <  \phi_{UV} \sim M_{UV}$ .
%%%%%%%%%%
\begin{center}
\begin{figure}
                \includegraphics[width=7cm,height=6cm,angle=0]{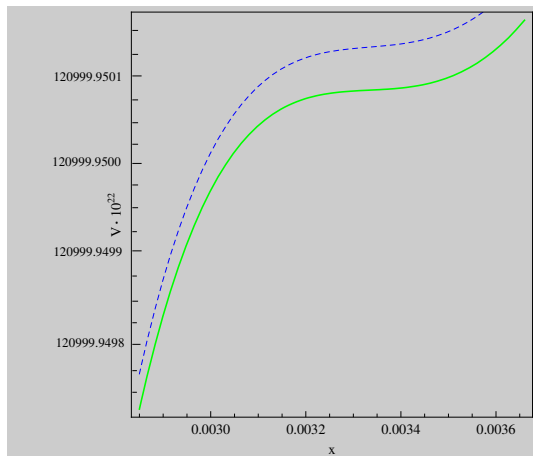}
 \caption{ The dashed line shows the plot of the effective potential given by (1) for  $ C_{3/2}= 0.006232,
 \alpha^{-1}= 2.11869,  {\cal D}=1.21\times 10^{-17} $ whereas the solid line  is the plot of the
 effective potential given by (3) for $C_1=10^{-7},C_{3/2}= 0.006232,\alpha^{-1}= 2.11869, C_2=10^{-6},  C_{5/2}=C_L=2\times 10^{-5},
   {\cal D}=1.21\times 10^{-17}$ }
 \label{pot1}
\end{figure}
\end{center}
%%%%

\begin{center}
\begin{figure}
\includegraphics[width=7cm,height=6cm,angle=0]{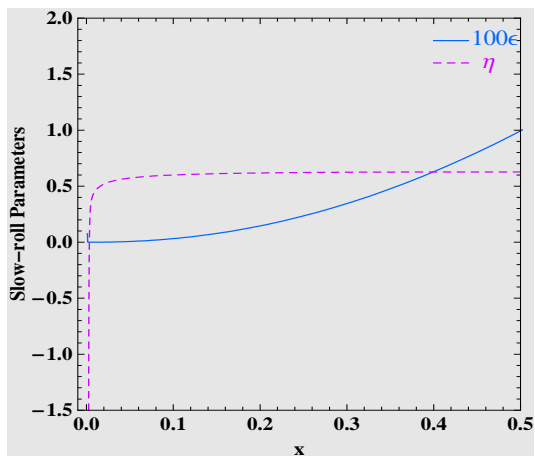}
 \caption{ Plot of the slow roll parameters $\epsilon$ $\&$ $\eta$  for the effective potential
 (3)  for the same set of parameters as in Fig. 1}
  \label{slp}
\end{figure}
\end{center}
%%%
The slow roll parameters for the generic field range are
\begin{eqnarray}
\label{epsilon1}
&&\epsilon= \frac{1}{2 \alpha^2}\left(\frac{{\cal V}_{,x}}{{\cal
V}(x)}\right)^2 \simeq \frac{1}{2 \alpha^2}\Big[
2(\frac{\alpha^2}{3}-C_2)x-C_1-\frac{3 C_{3/2}}{2} x^{1/2}
-\frac{5}{2} C_{5/2}x^{3/2}- L C_{L}x^{L-1}+\frac{3 {\cal D}}{4 \pi^2 \alpha^4 x^5}\Big ]^2 \\
&&\eta=\frac{1}{\alpha^2}\frac{{\cal V}_{,xx}}{{\cal V}(x)}\simeq \frac{1}{\alpha^2}\Big[
     2(\frac{\alpha^2}{3}-C_2)-\frac{3C_{3/2}}{4 x^{1/2}}-\frac{15}{4} C_{5/2}x^{1/2}- 2L  C_Lx^{L-2}-\frac{15{\cal
D}}{4 \pi^2 \alpha^4 x^6}\Big]
\label{eta1}
\end{eqnarray}
The COBE normalization and the spectral index are given by
\begin{equation}
\delta_H^2 \simeq \frac{1}{150 \pi^2M_P^4} \frac{V}{\epsilon}=
\frac{\alpha^4}{150 \pi^2}\frac{{\cal V}}{\epsilon}
\end{equation}
\begin{equation}
n_{s}=1+2\eta-6\epsilon
\end{equation}
In Fig.\ref{pot1}, we plot the effective potentials given by (1) and
(3) for a possible choice of parameters. It is clear from the figure
that the introduction of new terms in model (3), it is possible to
flaten the potential  in the field range of interest where inflation
takes place. This feature of the new potential effects the evolution
of the inflaton field and hence the inflation dynamics in a
considerable manner.

 We further note that the slow roll parameters satisfy
the relation $|\epsilon|<|\eta|$, in the case under consideration,
for the field range of interest as plotted in Fig.\ref{slp}.  Thus
it is sufficient to consider the parameter $\eta$ for discussing the
slow roll conditions and the end of inflation. Note that the field
potential should be monotonously increasing function of $x$ to
ensure realistic motion of D-brane, i.e., it should be allowed to
move towards the origin (tip of the throat) where the anti-D3-brane
is placed.

However, in general, the potential function can be monotonously
increasing,
 decreasing or even can acquire a minimum for $0<x<1$ depending upon the numerical values of the model parameters.
 Thus the requirement that it should be increasing  imposes constraints on the  parameters of the model.
 Moreover these constraints on parameters should allow us to have at least sixty e-foldings as well as
 should meet the experimental bounds on various cosmological observables.

As reported in \cite{ACPS}, the model described by (1)  exhibits
sensitivity with respect to the two parameters it has, namely,
$C_{3/2}$ and ${\cal D}$ and it extremely fine tuned with respect
the initial values of the inflaton field $x_i$. The monotonicity of
${\cal V}(x)$ can be ensured provided we require smaller and smaller
values of $C_{3/2}$. This is because we need to avoid the occurrence
of a minimum of the potential as we move towards the origin before
the effect of Coulomb term could become dominant.  For numerical
values of ${\cal D}^{1/4} \sim 10^{-4}$ required  for observational
constraints to be satisfied, we find numerical values of $C_{3/2}$
not only much smaller than one but also need  to be acutely fine
tuned. The field range viable for inflation turned out to be very
narrow resulting in number of e-foldings to be about 60 with the
initial value of the field $x_i$ being fine tuned typically to the
level of one part in $10^{-7}$. Any deviation of numerical values of
$x_i$ ($x_i=0.0033050$)\cite{ACPS} beyond the said accuracy makes
the inflaton hit the singularity before it could make the required
number of e-folds.  As a result, the inflationary scenario based
upon the effective potential (\ref{Spot}) becomes heavily
constrained.

%%%%%
\begin{center}
\begin{figure}
\includegraphics[width=7cm,height=6cm,angle=0]{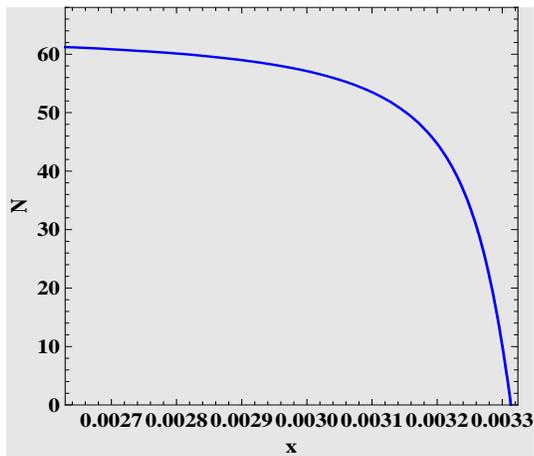}
 \caption{ Plot of the number of e-folds for the effective potential   (3)  versus
 $x$. The numerical values of model parameters is same  as in Fig 1}
  \label{Power1}
\end{figure}
\end{center}

%%%%%%

\begin{center}
\begin{figure}
\includegraphics[width=7cm,height=6cm,angle=0]{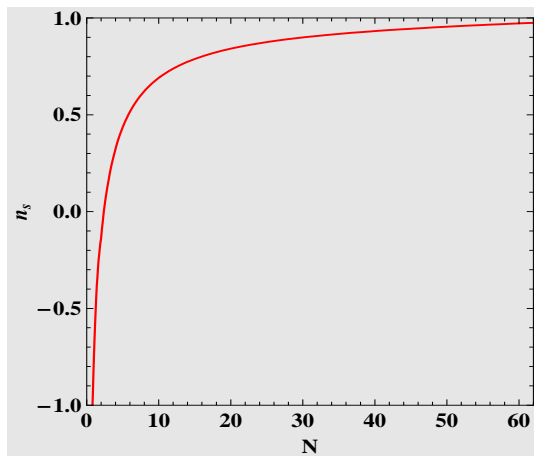}
 \caption{ Plot of the spectral index $n_S$ versus the field number of e-folds starting from the end of inflation for the effective potential   (3).   For the set of parameters as in Fig 1, $n_S$
reaches the observed value for $N\simeq 60$.}
 \label{Power2}
\end{figure}
\end{center}
%%%%%

To make the inflation model, at least a semi-realistic one and free
from severe fine tune tunings, we turn to the case of the improved
effective potential given by (\ref{Spot2}). As we will see, the fine
tuning of the
 parameters as well as that of the initial condition could be reduced. However, one might think that this
 is achieved at the cost of four new terms in the potential with corresponding
free parameters. This is indeed  true; nevertheless , it should  be
kept in mind that these new correction terms are not inspired by
phenomenological considerations but are rigorously derived using the
formalism of flux compactification and Gauge/Gravity correspondence.

 As pointed out by McAllister \cite{MAC}, if we restrict
to operators with dimension 4 for perturbation in the gauge theory
side, there are 324 terms contributing to inflation potential.
However, since the inflation is taking place for very small values
of the field ($x$ in our notation) and accompanied by parameters
which are also much less than one, terms involving higher powers of
$x$ can be safely neglected. The $x^2$ term actually plays the crucial
role since it affects the slow roll parameter ($\eta$). The terms
with powers less than 2 play an important role in solving the
eta-problem and maintaining the flatness of the potential to give at
least sixty e-foldings. We kept the term with irrational power
$(2\sqrt{7}-5/2)$ of the field for curiosity and to emphasize the fact that such
a term appears in the potential. Thus keeping the term $x^3$ in the
potential would not change our analysis and the result except for
allowing some more freedom with an extra parameter. The inclusion of
new terms extends the range of flatness allowing us to generate upto
70 or more efoldings without fine tuning of the initial position of
inflaton. The extra flat region is difficult to show on the plot
clearly. Our numerics clearly confirms it. The later is the major
contribution of the extra terms in the potential.

Our investigation shows that the numerical values of the new
constants are robust and can be varied over wide range with the
observational constraints satisfied. However, we should point out
that with many parameters in the potential, it is difficult to scan
the parameter space and put bounds on each of them. The best we have
achieved is to find a set of parameters by trial and error method to
satisfy the observational constraints coming from WMAP  seven years
data \cite{KAM2} and then to examine how robust the set is.  Table.I
and Table.II, represents
 such a collection of data points corresponding to different values of $\alpha^{-1}$.
%%%%%
\begin{table}[ht]\small
\hspace{-3cm}
  \begin{tabular}{|c|c|c|c|c|c|c|c|c|} \hline
 $C_{1}$  & $C_{3/2}$ & $\alpha^{-1}$ & $C_{2}$ & $C_{5/2}$ & $C_L$ & ${\cal D}$ & $n_{s}$ & $\delta^{2}_{H}$\\\hline
  $10^{-7}$& $623.2\times 10^{-5}$ & 2.11869     & $10^{-6}$ & $2\times 10^{-5}$  & $2\times 10^{-5}$ & $1.21\times10^{-17}$ & 0.976 &$2.38\times 10^{-9} $ \\\hline
  $1.9\times10^{-7}$& $623.2\times 10^{-5}$ & 2.11869 & $10^{-6}$ & $2\times 10^{-5}$  & $2\times 10^{-5}$ & $1.21\times10^{-17}$ & 0.950 &$2.45\times 10^{-9} $ \\\hline
  $10^{-7}$& $623.3\times 10^{-5}$  & 2.11869     & $10^{-6}$ & $2\times 10^{-5}$  & $2\times 10^{-5}$ & $1.21\times10^{-17}$ & 0.951 & $2.50\times 10^{-9} $ \\\hline
  $10^{-7}$& $623.2\times 10^{-5}$ & 2.11890     & $10^{-6}$ & $2\times 10^{-5}$  & $2\times 10^{-5}$ & $1.21\times10^{-17}$  & 0.952 & $2.43\times 10^{-9} $ \\\hline
  $10^{-7}$& $623.2\times 10^{-5}$ & 2.11869    & $10^{-10}$ & $2\times 10^{-5}$  & $2\times 10^{-5}$ & $1.21\times10^{-17}$ & 0.977 & $2.34\times 10^{-9} $ \\\hline
  $10^{-7}$& $623.2\times 10^{-5}$ & 2.11869 & $10^{-6}$ & $10^{-4}$  & $2\times 10^{-5}$ & $1.21\times10^{-17}$  & 0.976 & $2.38\times 10^{-9} $  \\\hline
  $10^{-7}$& $623.2\times 10^{-5}$ & 2.11869 & $10^{-6}$ & $2\times 10^{-5}$  & $10^{-10}$ & $1.21\times10^{-17}$  & 0.962 & $2.39\times 10^{-9} $ \\\hline
  $10^{-7}$& $623.2\times 10^{-5} $ & 2.11869     & $10^{-6}$ & $2\times 10^{-5}$  & $2\times 10^{-5}$ & $1.209\times10^{-17}$ & 0.963 & $2.37\times 10^{-9} $ \\\hline
  \end{tabular}
\caption{ A possible variation of model parameters consistent with
observational bounds: $\delta_{H}^{2}=(2.349-2.529)\times10^{-9}$
$\&$   $n_{s}=0.949-0.977(WMAP+BAO+H_{0})$(Tensor to scalar ratio of
perturbations is low in the model and does not impose constraints on
the model parameters)}
\end{table}
%%%

 %%%%%
\begin{table}[ht]\small
\hspace{-3cm}
  \begin{tabular}{|c|c|c|c|c|c|c|c|c|} \hline
 $C_{1}$  & $C_{3/2}$ & $\alpha^{-1}$ & $C_{2}$ & $C_{5/2}$ & $C_L$ & ${\cal D}$ & $n_{s}$ & $\delta^{2}_{H}$\\\hline
 $10^{-7}$& $3.63\times 10^{-3}$ & 3     & $10^{-4}$ & $2\times 10^{-5}$  & $10^{-3}$ & $10^{-17}$ & 0.977 &$2.35\times 10^{-9} $ \\\hline
$10^{-7}$& $3.62\times 10^{-3}$ & 3     & $10^{-5}$ & $2\times 10^{-5}$  & $10^{-3}$ & $10^{-17}$ & 0.965 &$2.3\times 10^{-9} $ \\\hline
 $10^{-9}$& $4.50\times 10^{-3}$ & 4    & $10^{-6}$ & $2\times 10^{-7}$  & $10^{-6}$ & $2\times 10^{-17}$ & 0.965 &$2.3\times 10^{-9} $ \\\hline
  $10^{-9}$& $4.50\times 10^{-3}$ & 4    & $10^{-5}$ & $2\times 10^{-6}$  & $10^{-6}$ & $2\times 10^{-17}$ & 0.96 &$2.34\times 10^{-9} $ \\\hline
  \end{tabular}
\caption{ A set of other possible variation of model parameters, showing the flexibility of $C_{3/2}$, consistent with observational bounds.}
\end{table}
%%%%%

We find that the effective potential  incorporating the additional
corrections is flatter over a generic range of field value for a set
of numerical values of model parameters. The later relaxes the fine
tuning of initial conditions for slow roll inflation and allows to
obtain the required number of sixty e-flodings  and even more, see
Fig.\ref{Power1}. In this case the initial value of the field $x_i$
needs to be fine tuned only  to the level of one part in $10^{-4}$.
In the Table I , we display a collection of sets of the numerical
values of the parameters that result from the numerical search
consistent with the observationally allowed range of values of
density perurbations $\delta_H^2 $, $n_S$ and tensor to scalar ratio
of perturbations. We vary  one parameter at a time, keeping the
other parameters fixed with numerical values as given . In Table
II, we present another viable set of model parameters corresponding
to numerical values of $\alpha$ different from those used in Table I.We
have confirmed that the new set of parameters can also give rise to
required number of e-folds and is consistent with observations. Note that
the fine tuning associated with $C_{3/2}$ improves by one order of
magnitude as compared to the earlier analysis based upon
(\ref{Spot}). In view of the above discussion, one might expect
enormous improvement with regard to COBE normalization ${\it a ~la}$
${\cal D}$. Actually, the improvement in this case is not beyond one
order of magnitude which is related to the fact that ${\cal D}$
appears in the effective potential in a non-trivial way i.e. besides
being an over all factor, it also appears in the Coulomb term.
Hence, any change in the numerical value of ${\cal D}$ crucially
effects the values of other parameters subject to the observational
constraints. In Fig.\ref{Power2}, we have shown the spectral index
versus the number of e-folds for
 a the same choice of parameters as used in Fig.1.

  \section{Conclusions}
 In this paper, we have investigated an inflationary model based upon an effective D-brane potential (\ref{Spot2})
that includes corrections to the potential arising from imaginary anti-self-dual fluxes encoded by terms containing coefficients
$C_1, C_2, C_{5/2},$ and $C_L$.  In\cite{ACPS}, where these terms were absent, it was
observed that the monotonocity of the potential imposes tough constraints on $C_{3/2}$. It needed to be fine
tuned to the level of one part in $10^{-7}$ for observational constraints to be satisfied. The potential was flat near the origin
for a very narrow field range and that too for only resticted values of model parameters. The constant ${\cal D}$ required heavy fine tuning
to satisfy the COBE constraint and the observational data on the spectral index $n_S$. What was worse,
the initial condition for the inflaton required fine tuning at the level of one part in $10^{-6}$ otherwise
the inflaton rolls to the region of instability before it completes 60 e-folds. The latter is related to
 very narrow field range of flatness of the potential (\ref{Spot}).
In the present model, there are more parameters (corresponding to the corrections due to  imaginary
anti-self-dual fluxes)  in the potential (\ref{Spot2}) which allow to increase the range of flatness of
the potential around $x=0$ thereby relaxing the fine tuning of initial conditions of the inflationary dynamics.
The modified D-brane potential can easily give rise to a large number of e-folfings without invoking much
fine tuning of initial conditions of the inflaton which is one of the major advantages of new correction terms
 in the effective D-brane potential. As for the  new constants, they do not involve much fine tuning for
  observational constraints to be satisfied. The fine tuning of model parameters $C_{3/2}$ and $ {\cal D}$ also
  improves by one order of magnitude respectively in the scenario based upon the corrected potential. Thus the
   inflationary scenario based upon the corrected potential performs much better than the earlier models
   of D-brane inflation. However, it should be noted that we have
   retained only five terms in the effective potential out of 324
   terms dictated by string theoretic considerations. It would be
   interesting to investigate the impact of other terms on the
   dynamics of inflaton. Secondly, our search of parameter space was
   based upon hit and trial method. The more sophisticated method
   based upon Monte-Carlo method could give rise to larger parameter
   space.

Last but not least, a viable inflation should be followed by a
successful reheating. Reheating in the scenario under consideration,
 could occur at the time of collision of D-brane with the ${\bar D}$-brane located at the tip of the throat which is beyond the
 regime of perturbative  string theoretic frame work used to obtain the effective potential (\ref{Spot2}). From phenomenological
 considerations, it looks quite plausible to implement here the instant reheating mechanism suitable to the class of models of
 non-oscillatory type. Since the new corrections to the D-brane potential are insignificant in the region where
 inflation ends, the preheating temperature is of the same order as obtained in case of the effective potential (1)\cite{PST}.

%%%%%%
\section*{ACKNOWLEDGMENTS}
 AD and SP thank the Centre for theoretical physics, Jamia Millia Islamia, New Delhi for hospitality. This work
 is partially supported by grant No. 2009/37/24 BRNS. MS thanks ICTP,Trieste, Italy for hospitality where a part of the work was done.

\end{document}